\documentclass[prd,aps,showpacs,secnumarabic]{revtex4}
\usepackage{graphicx}
\usepackage{subeqnar}
\begin{document}

\def\beqra{\begin{eqnarray}}
\def\eeqra{\end{eqnarray}}
\def\beqast{\begin{eqnarray*}}
\def\eeqast{\end{eqnarray*}}
\def\beq{\begin{equation}}
\def\eeq{\end{equation}}
\def\subeqn{\begin{subeqnarray}}
\def\endeqn{\end{subeqnarray}}
\def\ds{\displaystyle}
\def\ss{\scriptstyle}
\def\sb{\mbox{\rule{0pt}{11pt}}}
\parskip=4pt
\title{Comment on ``Evolution of a Quasi-Stationary State"}
\author{Duane A. Dicus$^1$, Wayne W. Repko$^2$, Roy F. Schwitters$^1$ and Todd M.
Tinsley$^1$ } \affiliation{ $^1$Center for Particle Physics, University of
Texas, Austin, TX 78712} \affiliation{$^2$Department of Physics and Astronomy,
Michigan State University, East Lansing, MI 48824}

\date{\today}

\begin{abstract}
Approximately forty years ago it was realized that the time development of
decaying systems might not be precisely exponential.  Rolf Winter (Phys. Rev.
{\bf 123}, 1503 (1961)) analyzed the simplest nontrivial system - a particle
tunneling out of a well formed by a wall and a delta-function.  He calculated
the probability current just outside the well and found irregular oscillations
on a short time scale followed by an exponential decrease followed by more
oscillations and finally by a decrease as a power of the time.  We have
reanalyzed this system, concentrating on the survival probability of the
particle in the well rather than the probability current, and find a different
short time behavior.
\end{abstract}
\pacs{03.65.-w, 03.65.Xp}
\maketitle

\section{Introduction}
Unstable quantum mechanical systems are generally described in terms of an
exponential decay law: the probability $P(t)$ that a given system remains
after a time $t$, called the survival probability, is given by
\begin{equation}\label{exp}
P(t) =e^{-\lambda t}\,,
\end{equation}
where $\lambda$ is a constant.  However it has been known since  Khalfin's
work \cite{1} in 1958 that decays could not be precisely exponential.  At very
long times $P(t)$ must decrease more slowly than Eq.\,(\ref{exp}) and, given
somewhat restrictive assumptions on the energy distribution of the system, the
derivative of $P(t)$ must vanish at $t=0$ \cite{2}.

In 1961 Winter \cite{3} gave a concrete example of nonexponential decay by
analyzing a simple one-dimensional tunneling problem.  He calculated the
probability current just outside a delta-function barrier to determine the
rate of decay.  Winter gave approximate analytic expressions for the wave
function, but to be precise he was forced to calculate the current numerically.
Given the state of computers in 1961 even to calculate the current was a
computational tour de force and he did not calculate the survival probability
itself.

Until recently, the direct experimental observation of departures from
exponential decay has proved elusive. Indeed, precise $\beta$ decay
experiments covering a range of decay times less than 10$^{-4}t_{1/2}$ in
$^{60}$Co and out to $45\,t_{1/2}$ in $^{56}$Mn showed no deviation from
exponential decay \cite{norman}. More recently, several experiments have
reported measurement-induced suppression in quantum systems at short times. One
\cite{4} observed the Zeno effect \cite{5}, which is a result of
$dP(t)/dt|_{t=0}$ being zero, in stimulated emission. Another \cite{kwhzk}
examined the effects of interaction-free measurements with photons. In the
past few years, the first experiments on an unstable quantum system
\cite{raiz1,raiz2} were performed. These experiments measure the time dependent
Landau-Zener tunneling of trapped sodium atoms. They observe nonexponential
decay directly and also see a Zeno effect and an anti-Zeno effect.

In view of the fact that some observations are now being made, we have
revisited Winter's calculation.  We believe it is basically correct. Here, we
simply extend it slightly by calculating the survival probability which is the
quantity of direct experimental interest.  Some of the nonexponential effects
in the probability current, such as large oscillations at short times, are
considerably damped in the survival probability. Other effects, such as large
oscillations during the transition from exponential to power law fall off at
large times, are not damped.  We turn now to the details.

\section{Winter's Model}

In this model, the potential consists of an infinite potential barrier at
$x=-a$ and a delta-function at $x=0$,
\begin{equation}
V(x) = \left\{\begin{array}{ccc}
        \infty & {\rm for} & x<-a \\[4pt]
       U\delta \,(x) & {\rm for} & x\geq-a\,
       \end{array}\right.\,,
\end{equation}
with $U>0$. The initial state is taken to be totally within the well formed by
the barrier and the delta-function
\begin{equation}\label{psi0}
\psi(x,0) = \left\{\begin{array}{ccc}
           (2/a)^{1/2} \sin (n\pi x/a) & {\rm for} & -a\leq x \leq 0 \\[4pt]
            0  & {\rm for} & x\,<\,-a\;\;{\rm or}\;\;x\,>\,0
            \end{array}\right.\,.
\end{equation}
To find the time dependence of $\psi$, we solve Schr\"odinger's equation for
the energy eigenstates $\phi_k(x)$. Taking into account the vanishing of the
wave function at $x=-a$, we can write
\begin{equation}
\phi_k (x) = \left\{\begin{array}{ccc}
             A\,\sin k(x+a) & {\rm for} & -a\leq x\leq 0 \\[4pt]
             Be^{ikx}+Ce^{-ikx} & {\rm for} & x>0
             \end{array}\right.\,,
\end{equation}
where $k=(2mE)^{1/2}$ with $m$ and $E$ denoting the mass and energy of the
particle in the well.  The boundary conditions at $x=0$ then relate the
constants $A$, $B$ and $C$ as
\begin{eqnarray}
B & = &\frac{1}{2}\left[\left(1-i\frac{G}{q}\right)\sin q - i\cos q\right]A \\
[4pt]
C & = &\frac{1}{2}\left[\left(1+i\frac{G}{q}\right)\sin q + i\cos
q\right]A\,,
\end{eqnarray}
where $G=2maU$ and $q=ka$. This then gives
\begin{equation}\label{eigen}
\phi_k(x) = A\left\{\begin{array}{ccc}
                   \sin(q(x/a+1))   & {\rm for}  & -a<x<0 \\[6pt]
                   \sin(q(x/a+1))+\frac{\ds\sb G}{\ds q}\sin(q)\sin(qx/a) & {\rm
                   for}  & x\geq 0
                   \end{array}\right.\,,
\end{equation}
with $A$ determined by normalization. We choose to normalize the $\phi_k(x)$ as
\begin{equation}\label{norm}
\int_{-a}^{\infty}dx\,\phi_{k'}^*(x)\phi_k(x) = \delta(k'-k)\,.
\end{equation}
In view of Eq.\,(\ref{eigen}), we have
\begin{eqnarray}\label{normint}
\int_{-a}^{\infty}dx\,\phi_{k'}^*(x)\phi_k(x) & = &\left\{
\left[1+\frac{G}{q}\sin q\cos q'+\frac{G}{q'}\sin q'\cos q+\frac{G^2}{qq'}\sin
q\sin q'\right] \int_0^{\infty}\!dx\sin k'x\sin kx \right.\nonumber \\
[6pt] && +\left.\left[\frac{G}{q}\int_0^{\infty}\!dx\cos k'x\sin
kx+\frac{G}{q'}\int_0^{\infty}\!dx\cos kx\sin k'x \right]\sin q'\sin
q\right\}|A|^2\,.
\end{eqnarray}
Using the integrals
\begin{eqnarray}
& &\int_0^{\infty}\!dx\,\sin k'x\sin kx = \frac{\pi}{2}[\delta(k'-k)-\delta(k'+k)]\,, \\
[4pt]
& &\frac{1}{k}\int_0^{\infty}\!dx\cos k'x\sin kx+\frac{1}{k'}\int_0^{\infty}\!dx\cos kx\sin k'x
= \pi^2\delta(k)\delta(k')\,,
\end{eqnarray}
and the fact that $k,k'>0$, Eq.\,(\ref{normint}) gives
\begin{equation}\label{A}
|A| = \sqrt{\frac{\ds 2}{\ds \pi}}\frac{q}{\sqrt{\ds q^2+qG\sin 2q+G^2\sin^2
q}}\,.
\end{equation}
Expanding $\psi(x,0)$, Eq.\,(\ref{psi0}), in terms of the $\phi_k(x)$ using
Eq.\,(\ref{A}), the expansion coefficients $C(k)$ are
\begin{equation}
C(k)=\frac{2n\sqrt{\ds\pi a}}{q^2-n^2\pi^2}\frac{q\sin q}{\sqrt{\ds q^2+qG\sin
2q +G^2\sin^2 q}}\,.
\end{equation}
The wave function $\psi(x,t)$ for $t>0$ can then be expanded as
\begin{eqnarray}\label{psiT}
\psi(x,t) &=& \int^\infty_0  \,dk\,C(k)\,\phi_k(x)e^{-i\frac{\sb k^2}{2m}t} \\[4pt]
&=& 2n\,\sqrt{\frac{\ds 2}{\ds a}}\int^\infty_0\,dq
\frac{e^{-iTq^2}}{q^2-n^2\pi^2}\;\frac{q\sin q\;\;f(x,q)}{q^2+Gq\sin 2q+
G^2\sin^2q}\label{psit}
\end{eqnarray}
where $T=t/2ma^2$. In terms of $\ell=x/a$, the function $f(x,q)$ is
\begin{equation}
f(x,q) = \left\{\begin{array}{ccc}
         q \sin (\ell+1)q & {\rm for} & -a\leq x\leq 0\\[4pt]
         q\sin(\ell+1)q+G\sin q\,\sin \ell q & {\rm for} & x>0
         \end{array}\right.\,.
\end{equation}

\section{Discussion}

Using the wave function of Eq.\,(\ref{psit}), Winter calculated the probability
current at the barrier, $j(x,T)$ at $x=0$,
\begin{equation}\label{curr}
j(0,T)=\frac{\hbar}{2mi}\;\left[\psi^*(0,T)\frac{d\psi(0,T)}{dx}-
\frac{d\psi^*(0,T)}{dx}\psi(0,T)\right]\,.
\end{equation}
We repeated this calculation and our results, shown in Fig.\,\ref{expcurr},
agree with his. In particular they show large oscillations at $T\leq 0.8$.
Like Winter, we used $G=6$ and $n=1$.

However the survival probability
\begin{equation}\label{surprob}
P(T) = \int^0_{-a}dx\;|\psi(x,T)|^2
\end{equation}
has only small oscillations, on the order of 5\% as shown in
Fig.\,\ref{expprob} and Fig.\,\ref{prob}. This is not too surprising since, by
the continuity equation, $P(T)$ and $j(0,T)$ are related by
\begin{equation}
P(T)-P(0)=-2ma^2\int^T_0 dT'\;j(0,T')
\end{equation}
and the effect of the integral, in this case, is to smooth out the
oscillations.

The source of the oscillations is the second denominator of Eq.\,(\ref{psit}),
which gives an infinite number of poles in the complex $q$-plane.  Closing the
contour in such a way as to pick up the contributions from these poles gives a
sum of exponentials with different energies and lifetimes
\begin{equation}\label{pole}
e^{-iTq^2}\rightarrow e^{-i\varepsilon_iT}\, e^{-T/2\tau_i}\,.
\end{equation}
Winter gives approximate analytic expressions for the pole positions as a
function of $G$ but these expressions are not accurate for $G\leq20$.  In
Table\,\ref{tab1} we show the pole positions for several values of $G$ as well
as the lifetime and energy from each pole. Since the first lifetime is
significantly larger, the other terms die out quickly but not before causing
the deviations shown in Figs. \ref{expprob} and \ref{prob}.  In fact the
difference of the first two energies gives a period of oscillation of 0.25
which agrees exactly with these figures.

Once these small short time oscillations die out, the slope determined
numerically and shown in Figs.\,\ref{prob} and \ref{prob3} agrees precisely
with that given by the first pole in Table\,\ref{tab1}.  This shows that there
are no additional contributions from the extra pieces of the contour required
to close it until we get to large times. At large times the factor
exp$(-iTq^2$) oscillates rapidly and the contribution to the integral is
negligible except for $q$ very small.  As a consequence, we can expand the
integrand in powers of $q$ and the integral in Eq.\,(\ref{psit}) goes as
\begin{equation}\label{qint}
\int_0^\infty\,dqe^{-iTq^2} q^2 \sim \frac{1}{T^{3/2}}
\end{equation}
More precisely, if we close the contour at an angle of $-45^{\circ}$ with the
real axis the extra contribution, beyond the poles discussed above, is
\begin{equation}\label{t32}
\psi(x,T)=\frac{(1+i)(\ell+1)}{2n\pi(G+1)^2 (\pi a)^{1/2}T^{3/2}}+{\cal
O}\left(\frac{1}{T^{5/2}}\right) ,~~-a\leq x\leq 0\,.
\end{equation}
So at very large times the survival probability goes as $1/T^3$. This agrees
with our numerical results as shown in Fig.\,\ref{prob3} for $T$ greater than
18.

As the time dependence changes from exponential to an inverse power Winter
found large oscillations in the probability current.  Indeed those also occur
in the survival probability as shown in Fig.\,\ref{prob3}.  They arise from the
interference between the time behavior shown in Eq.\,(\ref{qint}) or
Eq.\,(\ref{t32}), which have no oscillatory dependence, and that of
Eq.\,(\ref{pole}). The energy from Table\,\ref{tab1}, $\varepsilon = 7.59$,
gives a period of oscillation of 0.82 which agrees exactly with the numerical
results of Fig.\,\ref{prob3}.

\section{Summary}

For this simple example we can understand the decay through a barrier at all
times. At small time the survival probability has small oscillations because of
interference between exponentials with different lifetimes. This is followed
by an exponential region once all the exponentials except the one with the
longest lifetime decay away. Oscillations occur during the transition to the
$1/T^3$ behavior which holds at large time. All this is summarized in
Fig.\,\ref{prob3}. Except for the reduced magnitude of the small time
oscillations these results for the survival probability are essentially what
Winter found for the probability current.

We close with a few comments.

The size and duration of the small time oscillations are limited in this
system by the largest lifetime being significantly longer than the others.
However one could conceive of a situation where the longest two or more
lifetimes are similar in magnitude.  Then the oscillations could last longer
and be larger, perhaps even to the extent that there would be no single pure
exponential region. While it is easy enough to construct a $C(k)$ which will
give this behavior, it is not clear that such a function is associated with a
physically realizable system.

We define the survival probability as the probability the particle remains in
the well. Usually the survival probability is defined in terms of the survival
of the initial state
\begin{eqnarray}
P(T) & = & |\langle\psi(0)|\psi(T)\rangle|^2 \\[4pt]
     & = & \left|\int^0_{-a} dx\psi^*(x,0)\,\psi(x,T)\right|^2\,.\label{altsurprob}
\end{eqnarray}
Since Winter used the probability current just outside the barrier it seems
clear our definition agrees with his.  For this problem it makes very little
difference which definition is used. This is illustrated at short times for $G
= 6$ in Fig.\,\ref{compare}, where the definition Eq.\,(\ref{altsurprob}),
computed using the first two poles in Table \ref{tab1}, is compared with the
corresponding two pole evaluation of Eq.\,(\ref{surprob}). The two pole
approximation turns out to be quite accurate for $0<T\leq 2$. We have checked
that this result holds for other values of $G$.
For long times, where we can use Eq.\,(\ref{t32}), the ratio of
Eq.\,(\ref{surprob}) to Eq.\,({\ref{altsurprob}) is $\pi^2/6$. The combination
of the two pole approximation for the short time behavior of the survival
probabilities Eqs.\,(\ref{surprob}) and (\ref{altsurprob}) together with the
long time behavior given by Eq.\,(\ref{t32}) is shown in Fig.\,\ref{longt}.

Note that the wavefunction, Eq.\,(\ref{psiT}), must be equal to the initial
wavefunction, Eq.\,(\ref{psi0}), at $t=0$ and therefore the current at the
barrier, $x=0$, is zero at $t=0$.  Alternately, simply substitute the explicit
form for $\psi(x,t)$ into Eq.\,(\ref{curr}) and note that each surviving term
goes as $\sin(q^2T)$.  Thus
\begin{equation}
\frac{d}{dt}\;P(0) =-j(0,0) =0
\end{equation}
in agreement with more general arguments \cite{2}.

Since the small time behavior of $P(T)$ is larger than pure exponential (see
Fig.\,\ref{expprob}) the Zeno effect holds. The survival probability at a
particular time, say $T=1.0$, will be larger if intermediate measurements are
made.  For example, if we compare ten measurements at intervals of $T=0.1$ and
two measurements at intervals of $T=0.5$ with no additional measurements then,
using $P(T=0.1)=0.9119$, $P(T=0.5)=0.4754$ and $P(T=1.0)=0.2194$, we find
\begin{equation}
\left[P(T=0.1)\right]^{10} > \left[P(T=0.5)\right]^2 > P(T=1.0)\,.
\end{equation}
At very large times where $P(T)$ is again not exponential there is an
anti-Zeno effect - intermediate measurements make the survival probability much
{\it smaller} \cite{5}. For example
\begin{equation}
\left[P(T=15)\right]^{2} \ll P(T=30)\,.
\end{equation}

There can also be a short time anti-Zeno effect where the decay rate is
enhanced by repeated measurements at a time interval chosen to occur near the
minimum of the oscillations \cite{shl,kk1,kk2,fnp}. Observation of this is
reported in \cite{raiz2}. Such an effect is system dependent and may not occur
at all. Indeed, for our canonical case of $G=6$ it does not occur (note that
the in Fig.\,\ref{expprob} are always greater than unity). However, we have
checked that it does occur for other values of $G$, specifically $G=20$.

Finally it should be noted that there are deeper questions concerning quantum mechanics in a
time-asymmetric universe that we have not considered here \cite{7}.

\begin{acknowledgments}
One of us (D.A.D.) thanks Arno Bohm and Xerxes Tata for teaching him the basics
of nonexponential decay. This research was supported in part by the National
Science Foundation under grant PHY-0070443 and by the United States Department
of Energy under Contract No.DE-FG03-93ER40757.
\end{acknowledgments}

\begin{table}[h]
\centering
\begin{tabular}{lp{1.0in}p{1.0in}p{1.0in}p{1.0in}}
\toprule
& ${\rm Re}(q)$ & ${\rm Im}(q)$ & $\tau$ & $\varepsilon$ \\
\colrule
G=1 &&&&\\
\colrule
& 2.2986 & -0.76605 & 0.142 &4.70 \\
& 5.4340 & -1.19699 & 0.038 &28.10\\
& 8.5857 & -1.42451 & 0.020 &71.69\\
&11.7351 & -1.57997 & 0.013 &135.2\\
\colrule
G=6 &&&&\\
\colrule
& 2.7579 & -0.14043 & 0.646 &7.59 \\
& 5.7135 & -0.37015 & 0.118 &32.51 \\
& 8.7753 & -0.55535 & 0.051 &76.70 \\
&11.8767 & -0.69721 & 0.030 &140.6 \\
\colrule
G=20 &&&&\\
\colrule
& 2.9958 & -0.02054 & 4.063 &8.97 \\
& 6.0109 & -0.07438 & 0.559 &36.13 \\
& 9.0532 & -0.14565 & 0.190 &81.94 \\
&12.1204 & -0.22148 & 0.093 &146.9 \\
\botrule
\end{tabular}
\caption{Positions of the poles in the integrand of the expression for
$\psi(x,t)$. For each value of $G$ the positions of the first 4 poles are
given, as well as the lifetime, $\tau = -[4\,{\rm Re}(q) \;{\rm Im}(q)]^{-1}$,
and energy, $\varepsilon = ({\rm Re}(q))^2-({\rm Im}(q))^2$, for each pole.  In
the figures we use $G=6$.} \label{tab1}
\end{table}

\begin{figure}[h]
\centering\includegraphics[height=3.0in]{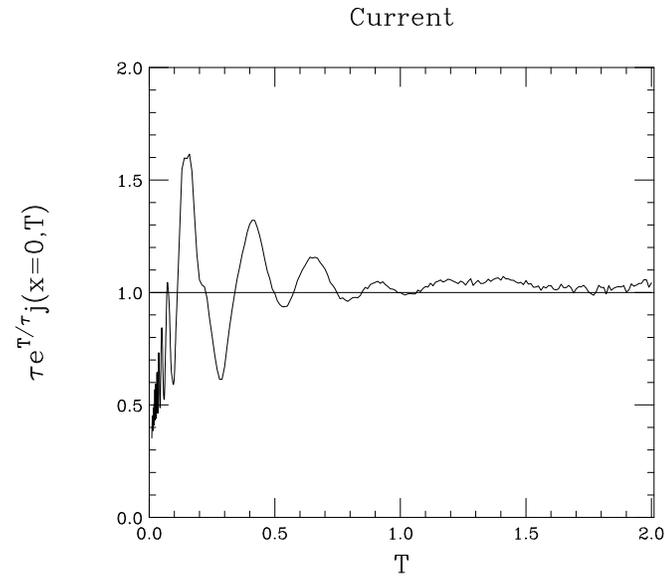} \caption{\footnotesize The
deviation of the probability current Eq.\,(\ref{curr}) from exponential
behavior at short times using $\tau=0.646$.\label{expcurr}}
\end{figure}

\begin{figure}[h]
\centering\includegraphics[height=3.0in]{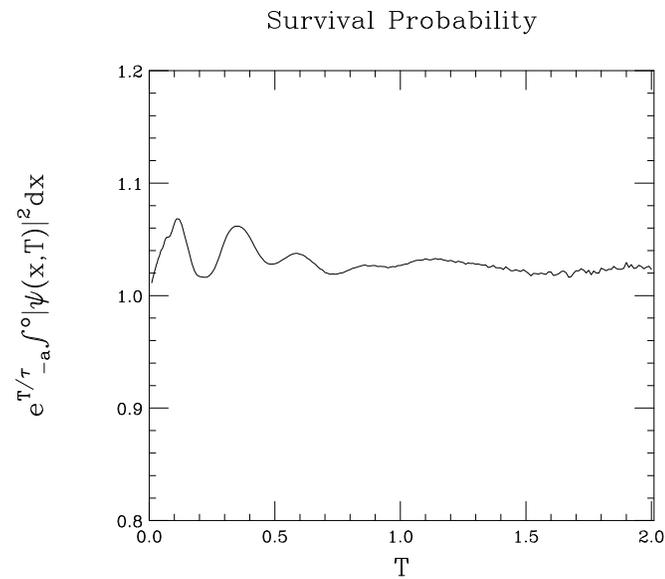} \caption{\footnotesize The
deviation of the survival probability Eq.\,(\ref{surprob}) from exponential
behavior at short times using $\tau=0.646$. \label{expprob}}
\end{figure}

\begin{figure}[h]
\centering\includegraphics[height=3.0in]{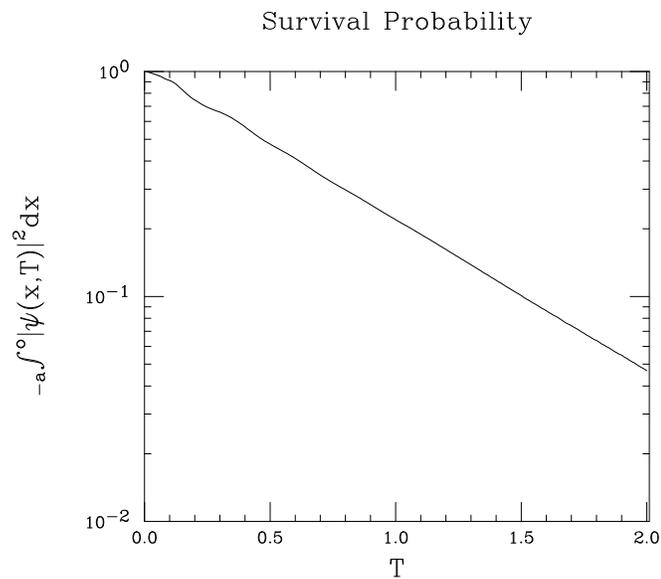} \caption{\footnotesize The
survival probability at short times. \label{prob}}
\end{figure}

\begin{figure}[h]
\centering\includegraphics[height=3.0in]{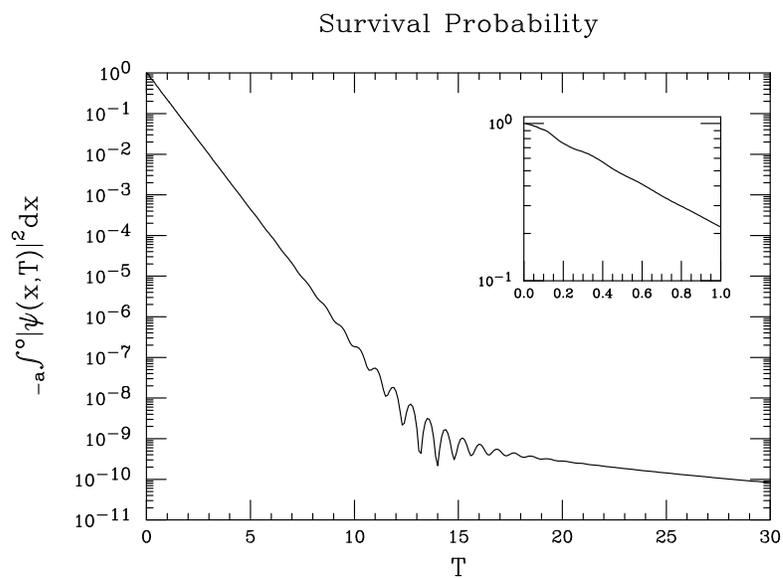} \caption{\footnotesize The
survival probability at all times. \label{prob3}}
\end{figure}

\begin{figure}[h]
\centering\includegraphics[height=3.0in]{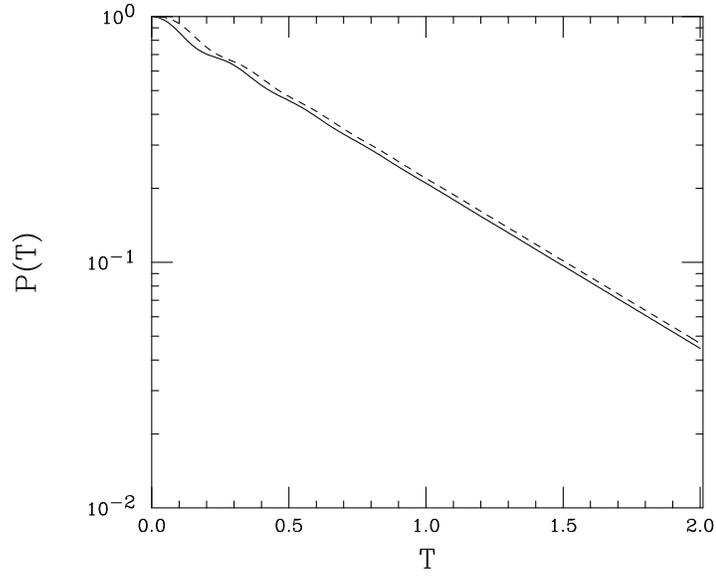} \caption{\footnotesize The
short time survival probability calculated from Eq.\,(\ref{altsurprob}) (solid
line) is compared with the survival probability calculated from
Eq.\,(\ref{surprob}) (dashed line) in the two pole approximation.
\label{compare}}
\end{figure}

\begin{figure}[h]
\centering\includegraphics[height=3.0in]{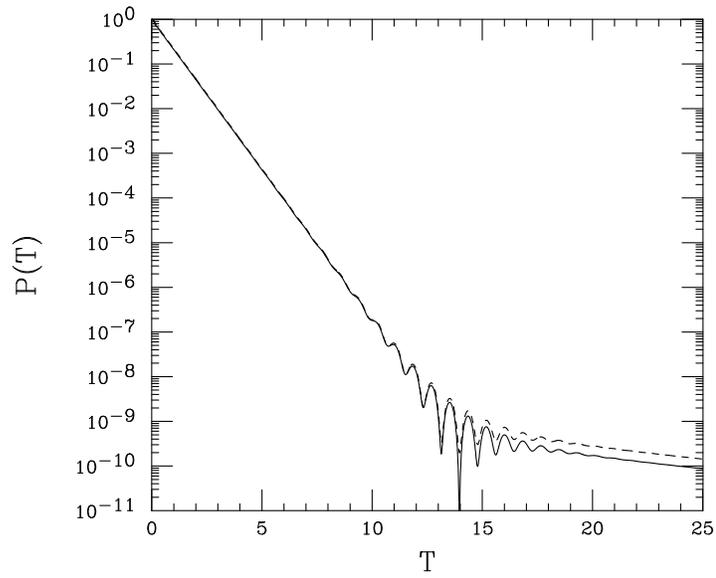} \caption{\footnotesize The
long time survival probability calculated from Eq.\,(\ref{altsurprob}) (solid
line) is compared with the survival probability calculated from
Eq.\,(\ref{surprob}) (dashed line) using the two pole approximation for $T<2$
and a combination of the leading pole and Eq.\,(\ref{t32}) for $T>2$.
\label{longt}}
\end{figure}

\end{document}